\documentclass[
10pt,
aps,
pra,
twocolumn,
groupedaddress,
superscriptaddress,
amsmath,
amssymb,
reprint,
longbibliography,
floatfix]{revtex4-2}

\usepackage{mathtools,amsmath,amsxtra}
\usepackage{amssymb}
\usepackage{physics}
\usepackage[english]{babel}
\usepackage{dsfont}
\usepackage{graphicx}% Include figure files
\usepackage{array}      % colonne personalizzate
\usepackage{float}      % posizionamento preciso
\usepackage{rotating}   % testo ruotato
\usepackage{makecell}   % celle comode (anche verticali)
\usepackage{dcolumn}% Align table columns on decimal point
\usepackage{bm}% bold math
\usepackage{bbold}
\usepackage{bigints}
\usepackage[dvipsnames]{xcolor}
\usepackage{tikz}
\usetikzlibrary{decorations.pathmorphing}
\usetikzlibrary{arrows.meta}
\usepackage{bm}
\usepackage{braket}
\usepackage[normalem]{ulem}
\usepackage{lipsum}
\usepackage{subcaption}
\usepackage{mathrsfs}
\usepackage{tabularx}
\begin{document}

\title{Raman relaxation in Yb(III) molecular qubits: non-trivial correlations between spin-phonon coupling and molecular structure}

\author{Giacomo Sansone}
\affiliation{Universit\`a di Parma, Dipartimento di Scienze Matematiche, Fisiche e Informatiche, I-43124 Parma, Italy}
\affiliation{INFN–Sezione di Milano-Bicocca, gruppo collegato di Parma, 43124 Parma, Italy}

\author{Lorenzo A. Mariano}
\affiliation{School of Physics, AMBER and CRANN Institute, Trinity College, Dublin 2, Ireland}

\author{Stefano Carretta}
\affiliation{Universit\`a di Parma, Dipartimento di Scienze Matematiche, Fisiche e Informatiche, I-43124 Parma, Italy}
\affiliation{INFN–Sezione di Milano-Bicocca, gruppo collegato di Parma, 43124 Parma, Italy}
\affiliation{Consorzio Interuniversitario Nazionale per la Scienza e Tecnologia dei Materiali (INSTM), I-50121 Firenze, Italy}

\author{Paolo Santini}
\affiliation{Universit\`a di Parma, Dipartimento di Scienze Matematiche, Fisiche e Informatiche, I-43124 Parma, Italy}
\affiliation{INFN–Sezione di Milano-Bicocca, gruppo collegato di Parma, 43124 Parma, Italy}
\affiliation{Consorzio Interuniversitario Nazionale per la Scienza e Tecnologia dei Materiali (INSTM), I-50121 Firenze, Italy}

\author{Alessandro Lunghi}
\email{lunghia@tcd.ie}
\affiliation{School of Physics, AMBER and CRANN Institute, Trinity College, Dublin 2, Ireland}

\begin{abstract}
The coordination complexes of Yb(III) exhibit some of the longest spin coherence times among 4f compounds, making them a promising platform for molecular quantum technologies. While spin-phonon relaxation remains a limiting factor for coherence times even at low temperature, its control through chemical design has the potential to push these spin qubits prototypes beyond current limits. With the aim of providing insights on how to chemically control spin-phonon relaxation, we here present a full ab initio study of spin–phonon dynamics for three Yb(III) molecules exhibiting minimal chemical differences, yet quantitatively different spin relaxation times. Results show that low-temperature relaxation is governed by Raman processes triggered by a small group of largely delocalized low-energy phonons. The analysis of these contributions highlights that the modulation of spin–phonon coupling by molecular structure modifications beyond the first coordination shell are highly non-trivial in nature and hard to rationalize in simple chemical terms. These findings call for a conceptual step change from the attempt to use simple magneto-structural correlations to interpret the effect of molecular structural modifications on spin-phonon relaxation, and present predictive first-principles frameworks as a potential driving force of future chemical design strategies.
\end{abstract}

\maketitle

\twocolumngrid

\section*{Introduction}
\label{sec:introduction}
Lanthanide(Ln)-based 4f Molecular Nanomagnets (MNM)  have been recently proposed as physical architectures for quantum information technologies due to their long relaxation and coherence times \cite{Pedersen2016,Chiesa2023,Chiesa2024, Hansen2024b}. A prototypical example is represented by the Yb(trensal) (\textbf{1}) molecule \cite{Pedersen2015}, whose $^{173}$Yb isotope has been vastly investigated and characterized over the last years. 
High resolution spectroscopies have revealed that within the $^2F_{7/2}$ ground term, the first excited electronic doublet lies nearly 500 cm$^{-1}$ above the fundamental one. This allows to treat Yb(trensal) as an effective electronic spin qubit ($S=1/2$) coupled through hyperfine interaction to a nuclear spin ($I=5/2$). With 12 addressable spin states, the Hilbert space can be expanded by encoding quantum information in a multi-level system (\textit{qudit}) \cite{Gottesman2001,Cafaro2012,Albert2020,Lim2025}, with the advantage of reducing the number of distinct physical units and two-body gates required for the implementation of many quantum operations.  
Transient nutation experiments with pulsed electron paramagnetic resonance (EPR) \cite{Pedersen2016} and nuclear magnetic resonance (NMR) \cite{Hussain2018} have demonstrated the possibility to generate and coherently manipulate pair superpositions of electro-nuclear states. 
In addition, magnetically diluted single crystals of Yb(trensal) have been placed on the inductors of lumped-element LC superconducting resonators, achieving high cooperativity coupling to all electronic and most nuclear spin transition, paving the way to the realization of qudit protocols in hybrid molecular spin--microwave photon processors \cite{Rollano2022}. 
Moreover, a recent work has reported the realization of the first prototype quantum simulator based on an ensemble of molecular qudits, exploiting as a quantum hardware a crystal containing isotopically enriched $^{173}$Yb(trensal), doped at 1\% into its diamagnetic [Lu(trensal)] isostructural analogue, and a radiofrequency broadband spectrometer \cite{Chicco2024}. A Quantum Fourier Transform algorithm could also be implemented, as recently shown in \cite{Rubin-Osanz2025}. \\
Despite all this progress, the interactions of the spin degrees of freedom with the external environment leads to fast decoherence above 10-20 K\cite{Pedersen2016}, which limits the temporal window within which it is possible to coherently control the spin states \cite{Ratini2025}. Above cryogenic temperatures ($T>4\,K$), the main limiting factor for coherence is the spin–phonon coupling, which sets the relaxation time $T_1$. In Yb(trensal), this mechanism completely suppresses coherent dynamics already above a few tens of Kelvin. 
\begin{figure*}[t]
    \centering
    \includegraphics[width=0.9\linewidth]{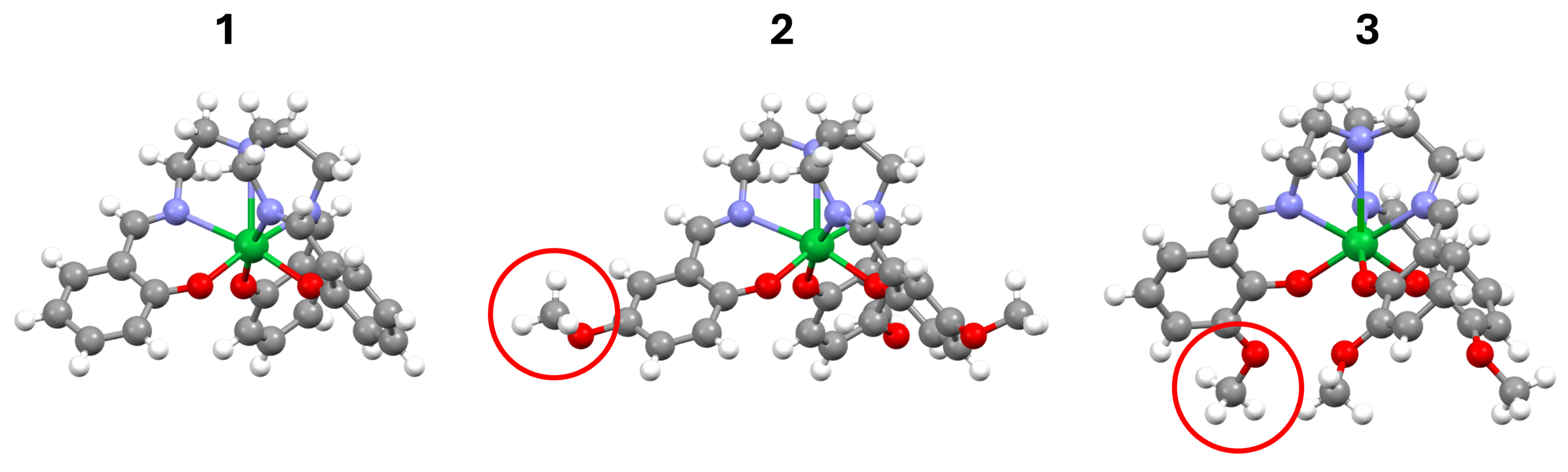}
    \caption{The molecular structure of Yb(trensal) (\textbf{1}), Yb(trenpvan) (\textbf{2}) and Yb(trenovan) (\textbf{3}). Color code: Yb = green, N = purple, O = red, C = grey, H = white. The red circles highlight the position of the methoxy groups that differentiate the three molecules.}
    \label{fig:tren-family}
\end{figure*}
Understanding how structural changes in molecular crystals affect spin dynamics is essential for the rational design of molecules with tailored properties. Substantial structural modifications that strongly perturb the spin states, the phonon spectrum, or both, are difficult to rationalize, as they complicate the disentanglement of the underlying mechanisms. A more effective starting point is to implement minimal structural modifications and systematically quantify their effects on the static and dynamic properties of the molecule. Recently, this idea has been applied precisely to Yb(trensal) \cite{Hansen2024a}. Two new derivative complexes, namely Yb(trenpvan) (\textbf{2}) and Yb(trenovan) (\textbf{3}), were indeed synthetized by modifying the trensal backbone. In particular, in each of the ligands constituting the tripodal scaffold, one hydrogen atom is substituted with a methoxy group (R--O--CH$_3$). Note that these compounds have identical chemical formulas and differ only in the placement of the methoxy group on the ligands. Fig.~\ref{fig:tren-family} shows the structure of \textbf{1} along with the ones of its two derivatives, \textbf{2} and \textbf{3}. A multi-technique study of the crystal field and spin-lattice relaxation of the two systems in the bulk phase was recently performed \cite{Hansen2024a}. That study revealed clear differences in the relaxation dynamics of these Yb coordination compounds despite their nearly identical spin states, thus indicating that even minor modifications of chemical groups in the second coordination sphere can affect the phonon spectrum and spin–phonon coupling. In the same work\cite{Hansen2024a}, infrared spectroscopy measurements also revealed differences in the optical phonon modes, suggesting them as a possible origin of the observed diversity in relaxation behavior. However, a clear understanding of the relaxation mechanism in Yb$^{3+}$ coordination complexes is still lacking, making it so far impossible to establish a direct link between chemical structure modifications and spin-phonon relaxation. \\ 

The picture that emerges from these studies underscores the need for a deeper understanding of the interplay between structural and electronic properties and relaxation dynamics. In this context, first-principles approaches to spin–phonon relaxation provide a natural and powerful framework to address this problem \cite{Lunghi2023}. Despite the growing interest in Yb(III) molecular qudits, a first-principles description of their spin–phonon relaxation is still lacking. Here, we address this issue through a fully ab initio study of spin–phonon relaxation in the family of nearly isostructural Yb(III) molecules introduced above, demonstrating how minimal chemical modifications can subtly tune Raman relaxation pathways. The article is organized as follows: first, we report the fundamental theory behind spin-phonon relaxation processes. Then, we illustrate the details of the numerical methods employed in the ab initio calculations of phonons, spin-phonon coupling and relaxation dynamics. Finally, we report and discuss the results of our simulations, both for the static properties and the relaxation dynamics of this family of qubits.

\section*{Theoretical background} \label{sec:theoretical_background}
The problem of  decoherence induced by spin-phonon coupling in the family of Yb(tren-) molecular qubits can be modelled through the following Hamiltonian
\begin{equation}   \hat{H}=\hat{H}_\mathrm{s}+\hat{H}_\mathrm{ph}+\hat{H}_\mathrm{{sph}} \:.
\end{equation}
Here, $\hat{H}_\mathrm{s}$ is the Hamiltonian describing the central system made of the $^{173}$Yb$^{3+}$ ion, $\hat{H}_\mathrm{ph}$ captures the contributions from the intrinsic degrees of freedom of the phonon bath, whereas $\hat{H}_\mathrm{sph}$ models the spin-phonon coupling. The central system Hamiltonian has the form
\begin{equation}\label{eq:spinHamiltonian}   \hat{H}_\mathrm{s}=\mathrm{\mu_B}\bm{B}\cdot \bm{g} \cdot \bm{\hat{J}} + \sum_{l=2,4,6} \sum_{m=-l}^lB_m^l\hat{O}_m^l(\bm{\hat{J}}) \:,
\end{equation}
where the first term describes the interaction between the ion spin $\bm{\hat{J}}$ and the external magnetic field $\bm{B}$ (being $\mathrm{\mu_B}$ and $\bm{g}$ the Bohr's magneton and the Landé tensor, respectively), while the second term describes the effective spin Hamiltonian, here modeled as a series of tesseral operators of order $l$ and component $m$, $\hat{O}_m^l$, and the related coefficients $B_m^l$. The phonon bath Hamiltonian is represented by the sum of quantum harmonic oscillators, 
\begin{equation}\label{eq:phononHamiltonian}   \hat{H}_\mathrm{ph}=\sum_{\bm{q}\alpha}\hbar \omega_{\bm{q}\alpha}\left(\hat{a}^\dagger_{\bm{q}\alpha}\hat{a}_{\bm{q}\alpha} + \frac{1}{2}\right) \:.
\end{equation}
Here, $\hbar$ is the reduced Planck constant, while $\omega_{\bm{q}\alpha}$ is the frequency of the phonon mode $Q_{\bm{q}\alpha}$ in branch $\alpha$ with reciprocal lattice vector $\bm{q}$. Finally,  $\hat{a}^\dagger_{\bm{q}\alpha}$ and $\hat{a}_{\bm{q}\alpha}$ are the phonon-specific creation and annihilation operators, respectively. From now on, we will drop the subscript $\bm{q}$ since our analysis will pertain only phonon modes at the $\Gamma$-point, i.e. $\bm{q}\equiv$ (0,0,0). 

The spin-phonon coupling Hamiltonian $\hat{H}_\mathrm{sph}$ allows us to describe the modulation effect of the lattice dynamics on the spin Hamiltonian coefficients, i.e. the parameters $B_m^l$ in Equation~\eqref{eq:spinHamiltonian}. Under the assumption of a weak coupling regime between the electronic spin and phonons, we can expand the spin Hamiltonian parameters in a Taylor series with respect to the set of vibrational mode coordinates, $\{Q_{ \alpha}\}$, around the equilibrium position $\{Q_{ \alpha}=0\}$,
\begin{equation}\label{eq:taylorExpansion}
    B_m^l\left(\{Q_{ \alpha}\}\right) = B_{m}^l(0) + \sum_{ \alpha} \left(\frac{\partial B_m^l}{\partial Q_{ \alpha}}\right)_{0} Q_{ \alpha} + \dots \:.
\end{equation}
Here, the $B_m^l(0)$ are the equilibrium spin Hamiltonian parameters appearing in Equation~\eqref{eq:spinHamiltonian}, while the first derivatives are the linear spin-phonon coupling coefficients. The subscript $0$ indicates that these derivatives are evaluated at the crystal equilibrium positions. By promoting the coordinates $Q_{ \alpha}$ to quantum operators $\hat{Q}_{ \alpha}$ we can then define the spin-phonon coupling Hamiltonian as
\begin{align}
    &\hat{H}_\mathrm{sph} = \sum_{l=2,4,6}\sum_{m=-l}^l\sum_{ \alpha} \left(\frac{\partial B_m^l}{\partial Q_{ \alpha}}\right)_{0} \hat{Q}_{ \alpha} \hat{O}_m^l \:. \label{eq:sphHamiltonian}
\end{align}

Once the eigenstates $\ket{a}, \, \ket{b}, \, \dots$ of the spin Hamiltonian and the corresponding eigenvalues $E_a, \, E_b, \, \dots$ have been obtained, it is possible to evaluate the transition rates between different spin states, $R_{ba}$, and use them to simulate the spin dynamics through time-local quantum master equations. 
In the following, we will consider contributions to relaxation coming from both one- and two-phonon processes.
Starting from one-phonon processes, the transition rate among two generic spin states, $R_{ba}^\mathrm{1-ph}$, reads
\begin{equation}\label{eq:R_1ph}
   R_{ba}^\mathrm{1-ph} = \frac{2\pi}{\hbar^2}\sum_{\alpha} \abs{V_{ba}^\alpha }^2G^\mathrm{1-ph}(\omega_{ba},\omega_\alpha),   
\end{equation}
where 
\begin{equation}
    V_{ba}^\alpha= \sum_{ml} \left(\frac{\partial B_m^l}{\partial Q_{\alpha}} \right) \bra{b} \hat{O}_m^l \ket{a}\:.
\end{equation}
Here, the term $( \partial B_m^l / \partial Q_\alpha )$ quantifies the intensity of the coupling between the spin and the phonon mode $Q_\alpha$, whereas $\hbar\omega_{ba} = E_b-E_a$ defines the energy gap between the two states involved in the spin transition.  
The function $G^\mathrm{1-ph}$ arises from the evaluation of the matrix element of the normal mode operator $Q_\alpha$ between two phonon eigenstates and reads
\begin{equation}\label{eq:G_1ph}
    G^\mathrm{1-ph}(\omega,\omega_\alpha) = \delta(\omega-\omega_\alpha)\bar{n}_{\alpha} + \delta(\omega+\omega_\alpha)(\bar{n}_\alpha+1)\:,
\end{equation}
where $\delta(\omega \pm \omega_\alpha)$ is the Dirac's delta function that selects phonons resonant with the spin gap in the absorption and emission processes, and $\bar{n}_{\alpha} = \left( e^{\hbar\omega_\alpha/\mathrm{k_B}T} -1 \right)^{-1}$ is the Bose-Einstein distribution accounting for the phonons thermal population, $\mathrm{k_B}$ being the Boltzmann constant. Two-phonon processes provide an alternative path to relaxation. For instance, the simultaneous absorption and emission of two phonons can be written as
\begin{equation}\label{eq:R_2ph}
    R_{ba}^\mathrm{2-ph} = \frac{2\pi}{\hbar^2} \sum_{\alpha\beta}\abs{T_{ba}^{\alpha\beta,+}+T_{ba}^{\beta\alpha,-}}^2 G^\mathrm{2-ph}(\omega_{ba},\omega_\alpha,\omega_\beta).
\end{equation}
Here, the terms $T_{ba}^{\alpha\beta,\pm}$ are defined as
\begin{equation}
    T_{ba}^{\alpha\beta,\pm} = \sum_c \frac{ V_{bc}^{\alpha}V_{ca}^{\beta}}{E_c-E_a\pm\hbar\omega_\beta}   \:, 
\end{equation}
whereas the function $G^\mathrm{2-ph}$ acts in a similar way to $G^\mathrm{1-ph}$ introduced in Equation~\eqref{eq:G_1ph} for one-phonon processes, imposing the energy conservation during a spin transition, and accounting for the thermal population of the states (Bose-Einstein distribution):
\begin{equation}\label{eq:G_2ph}
    G^\mathrm{2-ph}(\omega,\omega_\alpha,\omega_\beta)=\delta(\omega+\omega_\alpha-\omega_\beta)(\bar{n}_\alpha+1)\bar{n}_\beta 
\end{equation}\\

These two processes have been shown to be able to induce spin relaxation in coordination complexes of f-ions through Orbach and Raman-I relaxation mechanisms, respectively \cite{Blagg2013,Chiesa2020,Briganti2021,Lunghi2022}. We note that we do not consider here a second source of two-phonon processes, which originate from the second-order term in the expansion of the crystal-field parameters (Equation~\eqref{eq:taylorExpansion}), and lead to what is sometimes referred to as Raman-II relaxation \cite{Lunghi2022}. These contributions would lead to a spin relaxation mechanism that, like the direct one, is suppressed by the Kramers nature of the Yb ground-state doublet. This is little affected by the applied magnetic field, which is too small to induce appreciable mixing with excited states given their large excitation gaps shown in Fig.~\ref{fig:spectra}.

The relaxation rates computed through \eqref{eq:R_1ph} and \eqref{eq:R_2ph} allow to build a transition rate matrix $R_{ba}$, which contains the whole information related to the dynamics of the population (diagonal) terms of the density matrix, namely
\begin{equation}\label{eq:Redfield}
    \frac{d\rho_{bb}(t)}{dt} = \sum_{cd}R_{ba}\rho_{aa}(t)\:.
\end{equation}
A generalization of these equations to the full density matrix, thus including the effect of phonons on coherence terms, has recently been proposed \cite{Lunghi2025}, but here we will only be concerned with the study of population terms, which are decoupled from coherences in the absence of spectral degeneracies\cite{Lunghi2022}. 

\section*{Computational Methods} \label{sec:methods}
The cell and geometry optimization process and the simulations of $\Gamma$-point phonons for the unit cells of \textbf{1}, \textbf{2} and \textbf{3} were performed through periodic density functional theory (pDFT) calculations using the \textit{CP2K} package \cite{cp2k2020}. The original X-ray crystallographic structure was used as a starting point. Cell optimization was performed employing tight convergence criteria of 10$^{-7}$ a.u. for the forces and a Self Consistent Field (SCF) convergence criteria of 10$^{-10}$ a.u. for the energy. A plane wave cutoff of 1000 Ry, DZVP-MOLOPT Gaussian basis sets, and Goedecker--Teter--Hutter pseudopotentials \cite{Goedecker1996} were employed for all atoms. The Perdew-Burke-Ernzerhof (PBE) \cite{Perdew1996} functional was used, together with Grimme's D3 dispersion correction to properly describe van der Waals interactions in the molecular crystal. 
Phonons modes $Q_\alpha$ and the corresponding frequencies $\omega_\alpha$ were computed with a two-step numerical differentiation of forces and step 0.01 \AA. 
By means of the \textit{ORCA} (v.\,5) quantum chemistry package \cite{ORCA},  electronic structure and magnetic properties were computed on the optimized structures. The DKH-def2-TZVPP basis set was used for all the atoms, except Yb, for which SARC2-DKH-QZVP was employed. Douglas-Kroll-Hess (DKH) scalar relativistic correction to the electronic Hamiltonian was employed \cite{Douglas1974,Hess1986}. Picture-change effects were included in the calculation. 
Multireference calculations were performed using Complete Active Space Self Consistent Field (CASSCF). The active space used to build the CASSCF wavefunction is (13,7), i.e., thirteen electrons in seven 4f-orbitals.
The n-electron valence state perturbation theory at second order (NEVPT2 \cite{Angeli2001}) was applied to include corrections on the three systems. The spin-phonon coupling coefficients $(\partial B_m^l/\partial Q_\alpha)$ were calculated by recurring to two-step numerical differentiation of the crystal field parameters $B_m^l$ using a step of 0.01 \AA for each molecular degree of freedom. Their value was then retrieved through the interpolation with a second-order polynomial function, keeping the resulting linear coefficients.  
The Zeeman term for the static Hamiltonian in \ref{eq:spinHamiltonian} was built assuming an external static magnetic field of 0.05 T, oriented along the $\mathbf{\hat{z}}$ direction of the external, Cartesian reference frame. A rotation to align the magnetic field with the terms of the Hamiltonian expressed in the molecular reference frame was also applied by specifying the Euler's angles that define the rotation. Dirac’s delta functions appearing in the rate equations (see Equations~\eqref{eq:G_1ph} and \eqref{eq:G_2ph}) were smeared out with a Gaussian function with $\sigma$=10 cm$^{-1}$. Second and fourth-order time-local quantum master equations (Equations \ref{eq:R_1ph} and \ref{eq:R_2ph}), were used to simulate one- and two-phonon relaxation processes, respectively. In practice, $T_1$ is extracted from the matrix $R_{ba}$ of Equation \ref{eq:Redfield} as the first non-null eigenvalue. The software MolForge was used for these simulations \cite{MolForge2024,Lunghi2022,Lunghi2019,Lunghi2020}.

\section*{Results} 
\label{sec:resultsAndDiscussions}

\begin{figure}
    \centering
    \includegraphics[width=0.96\linewidth]{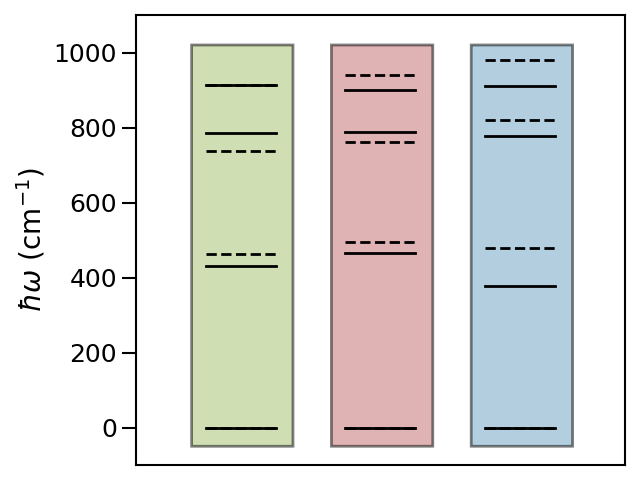}
    \caption{Energy spectra of the ground term $^2F_{7/2}$ for the three systems. The solid lines represent the energy levels retrieved from NEVPT2 calculations, whereas the dashed lines refer to the experimental spectroscopic values\cite{Pedersen2015,Kragskow2022,Hansen2024a}.}
    \label{fig:spectra}
\end{figure}

The energy spectra related to the spectroscopic ground term of Yb$^{3+}$, $^2F_{7/2}$, consist of four Kramers Doublets (KDs). The results of their ab initio calculation are summarized and compared to experimental measurements \cite{Pedersen2015,Kragskow2022,Hansen2024a} in Fig.~\ref{fig:spectra}. We observe excellent agreement between the calculated and experimental values, with a maximum deviation of $\sim 10\%$ for the highest excited doublet of Yb(trenovan). Table~\ref{tab:gtensor} summarizes the results related to the \textit{g}-tensor for the ground state doublet. With respect to the experimental values, the theoretical results tend to overestimate the \textit{g}-shift.
This is a feature that has been already observed in previous works on Yb$^{3+}$ specifically and on Ln-based complexes in general \cite{Kragskow2022,Ungur2017}.

\begin{table}
\centering

\begin{tabular}{c|c|c|c|c}
%  | >{\hsize=1.4\hsize\centering\arraybackslash}X  % Questa è il 40% più larga
%  | >{\hsize=1.0\hsize\centering\arraybackslash}X  % Standard
%  | >{\hsize=0.86\hsize\centering\arraybackslash}X % Più stretta
%  | >{\hsize=0.86\hsize\centering\arraybackslash}X % Più stretta
%  | >{\hsize=0.86\hsize\centering\arraybackslash}X |% Più stretta
%}
    \hline
    \hline
    Molecule & Method & $g_{xx}$ & $g_{yy}$ & $g_{zz}$ \\
    \hline
    \hline
    Yb(tren\textbf{sal}) &  Sim. & 2.68 & 2.73 & 4.88 \\
    Yb(tren\textbf{sal}) & Exp.  & 2.9& 2.9& 4.3 \\
    \hline
    Yb(tren\textbf{pvan}) & Sim.  & 2.51& 2.90& 4.91 \\
    Yb(tren\textbf{pvan}) & Exp & 2.96& 2.96& 4.16 \\
    \hline 
    Yb(tren\textbf{ovan}) & Sim.  & 2.46& 2.61& 5.24 \\
    Yb(tren\textbf{ovan}) & Exp.  & 2.65& 2.65& 4.89 \\
    \hline
    \hline
\end{tabular}
\caption{Comparison between simulated and experimental $g$-tensor values for the ground doublet of the three systems.}
\label{tab:gtensor}
\end{table}

\begin{figure*}[ht]
    \centering

    % --- Riga
    \begin{subfigure}{0.32\textwidth}
        \centering
        \includegraphics[width=\textwidth]{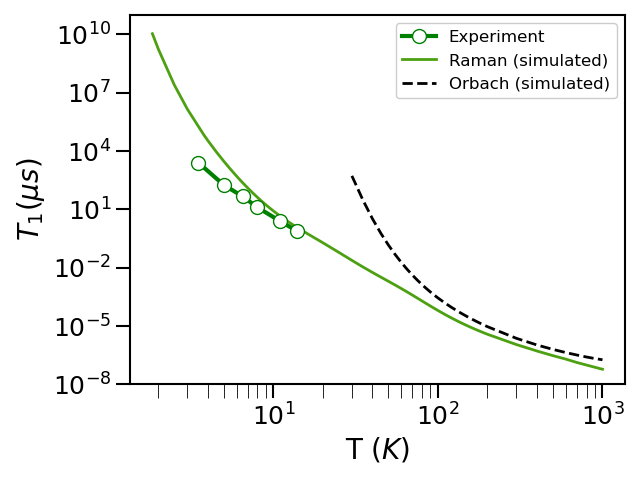}
        \caption{Yb(trensal) relaxation}
    \end{subfigure}
    \begin{subfigure}{0.32\textwidth}
        \centering
        \includegraphics[width=\textwidth]{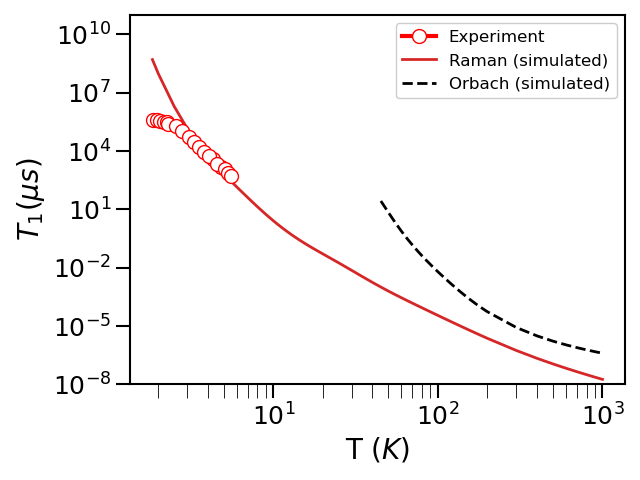}
        \caption{Yb(trenpvan) relaxation}
    \end{subfigure}
    \begin{subfigure}{0.32\textwidth}
        \centering
        \includegraphics[width=\textwidth]{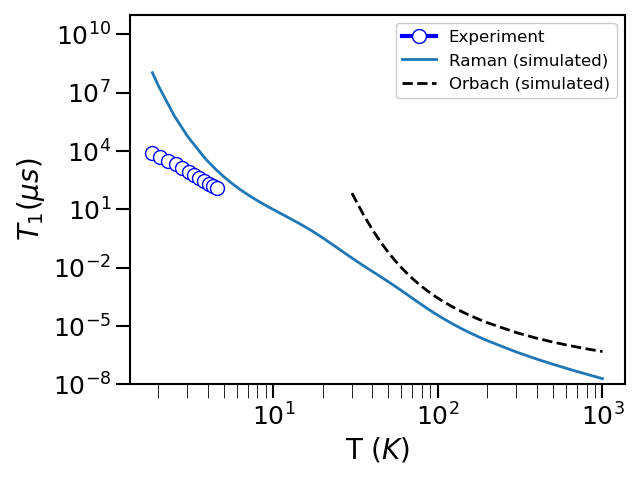}
        \caption{Yb(trenovan) relaxation}
    \end{subfigure}

    \vspace{1cm}    
    % --- Prima riga ---
    \begin{subfigure}{0.32\textwidth}
        \centering
        \includegraphics[width=\textwidth]{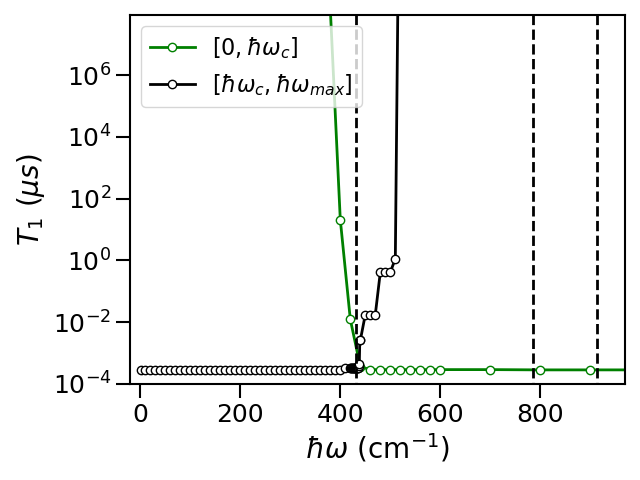}
        \caption{Yb(trensal) - Orbach}
    \end{subfigure}
    \begin{subfigure}{0.32\textwidth}
        \centering
        \includegraphics[width=\textwidth]{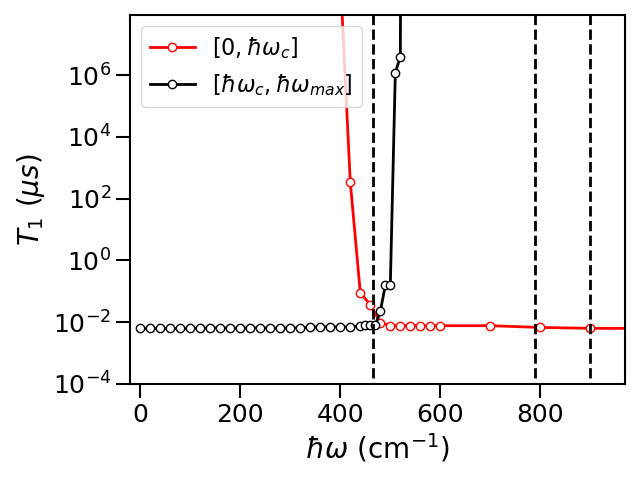}
        \caption{Yb(trenpvan) - Orbach}
    \end{subfigure}
    \begin{subfigure}{0.32\textwidth}
        \centering
        \includegraphics[width=\textwidth]{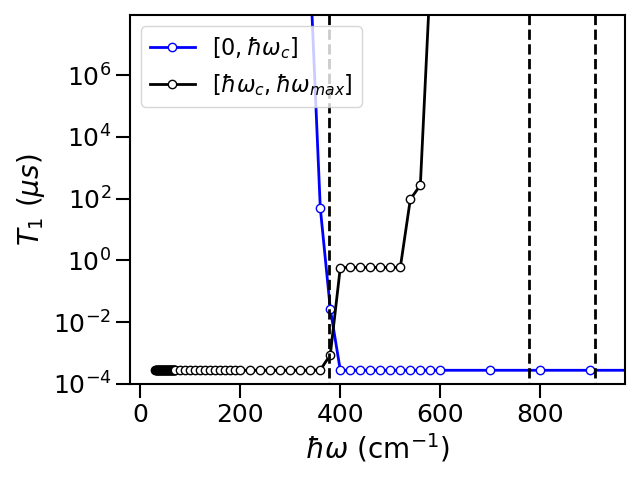}
        \caption{Yb(trenovan) - Orbach}
    \end{subfigure}

    \vspace{1cm}
    % --- Seconda riga ---
        \begin{subfigure}{0.32\textwidth}
        \centering
        \includegraphics[width=\textwidth]{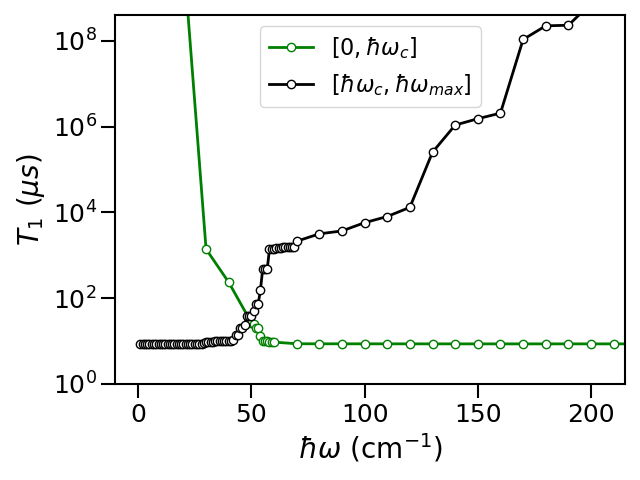}
        \caption{Yb(trensal) - Raman}
    \end{subfigure}
    \begin{subfigure}{0.32\textwidth}
        \centering
        \includegraphics[width=\textwidth]{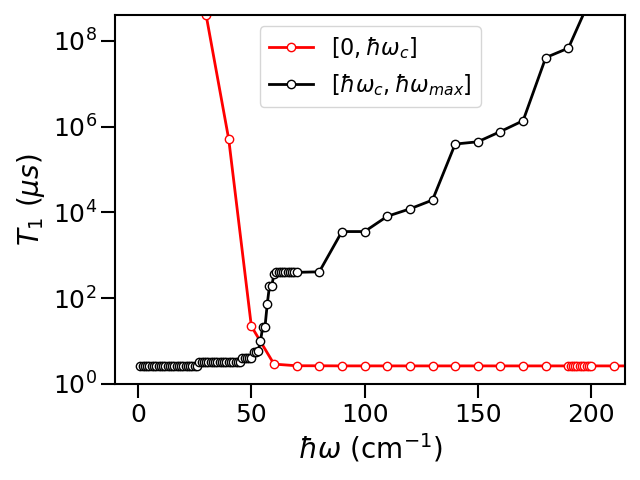}
        \caption{Yb(trenpvan) - Raman}
    \end{subfigure}
    \begin{subfigure}{0.32\textwidth}
        \centering
        \includegraphics[width=\textwidth]{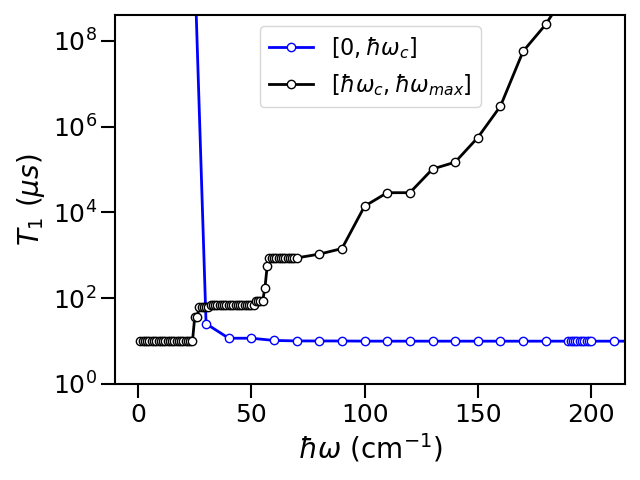}
        \caption{Yb(trenovan) - Raman}
    \end{subfigure}

    \caption{Relaxation time as a function of temperature for Yb(trensal) (a), Yb(trenpvan) (b) and Yb(trenovan) (c). Simulated Raman (solid line) and Orbach (dashed) contributions are reported. Experimental data (hollow circles) are taken from Ref.~\cite{Bode2023,Hansen2024a}. Panels d-f and g-i report the results of the energy cut-off analysis for each system, both in the case of Orbach relaxation (panels d-f, $T=100 \, K$) and Raman relaxation (panels g-i, $T= \, 10$~K), respectively. Colored curves report the relaxation time $T_1$ evaluated by considering phonons in the energy interval $[0,\omega_c]$, where $\omega_c$ is the cut-off frequency. The black curves report the same simulation for phonons in the interval $[\omega_\mathrm{c},\omega_\mathrm{max}]$, where $\omega_{max}$ is the fixed, maximum frequency of the phonons considered. The vertical, dashed lines mark the energy levels of the KD.}
    \label{fig:relaxation_and_cutoff_analysis}
\end{figure*}

The spin-phonon relaxation time $T_1$ was computed for the three systems as a function of temperature. The results are compared with experimental data in Fig.~\ref{fig:relaxation_and_cutoff_analysis}a-c.
Data related to \textbf{1} were taken from Bode et al.~\cite{Bode2023}, where pulsed EPR measurements were performed on a single-crystal of diamagnetic Lu(trensal) at 1\% Yb doping (Lu$_{0.99}$Yb$_{0.01}$(trensal)), with an external field $B_0=0.2012$~T oriented along the $C_3$ crystallographic axis. Data for \textbf{2} and \textbf{3} were taken from Hansen et al.~\cite{Hansen2024a}, and were obtained through dynamic magnetic susceptibility measurements on  diluted Lu$_{0.99}$Yb$_{0.01}$(trenpvan) and Lu$_{0.99}$Yb$_{0.01}$(trenovan), under an external field $B_0=0.05$~T also oriented along the $C_3$ axis. The calculations were performed assuming for all the three systems an external magnetic field $B_0 = 0.05$ T, oriented along the $C_3$ crystallographic axis. In the case of \textbf{1}, the simulation was performed also by setting $B_0=0.2012$ to perfectly match the experimental conditions. However, no appreciable dependence of $T_1$ on the field was noticed within the range of such small magnetic fields.\\
Results show that Orbach relaxation is negligible at low temperatures, as expected from the large crystal-field gaps (see Fig.~\ref{fig:spectra}), and from its reliance on real, thermally activated transitions from the ground to excited doublets. Thus, two-phonon processes alone determine (Raman) relaxation in the experimentally accessible temperature window. For \textbf{1} and \textbf{2} the calculations are in good agreement with experimental data in upper range of measured temperatures, whereas $T_1$ is overestimated at lower temperatures. This is not unexpected, when spin-phonon relaxation becomes very slow different minor relaxation mechanisms emerge and eventually set $T_1$ (e.g., quantum tunneling, dipolar interactions).
Unfortunately, in the case of \textbf{3}, we lack information for $T_1$ at temperatures higher than 5~K ($1/T<0.2\,K^{-1}$), making the simulated $T_1$ overestimated over the whole range of the available measurements.
It must be stressed that the differences in relaxation behavior among the three variants presented in Figure \ref{fig:relaxation_and_cutoff_analysis} are entirely ascribable to the spin-lattice dynamics. We quantitatively demonstrate this point by noting that no substantial impact on the relaxation dynamics of \textbf{1} is observed by interchanging its static crystal field Hamiltonian with the one of \textbf{2} or \textbf{3}, in agreement with the high level of similarity among the electronic spectra in Figure \ref{fig:spectra}.

To get a deeper insight into the relaxation mechanism and identify the vibrational modes that most strongly influence the spin dynamics, we calculate $T_1$ by selectively including phonon modes within specific energy windows. First, we  set the lower bound of the energy window at 0 cm$^{-1}$ while the upper bound $\omega_\mathrm{c}$ is progressively increased; at each step, $T_1$ is evaluated by including  phonons with energies in the interval $[0,\omega_\mathrm{c}]$. We  refer to this procedure as \textit{high cut-off analysis}. In a second approach, we fix the upper bound $\omega_\mathrm{max}$ of the energy window, and the lower bound $\omega_\mathrm{c}$ is progressively increased from 0 cm$^{-1}$ up to $\omega_\mathrm{max}$; again, at each step $T_1$ is calculated considering the contribution of phonons with energy in the interval $[\omega_\mathrm{c},\omega_\mathrm{max}]$ (\textit{low cut-off analysis}). Results  are summarized in Fig.~\ref{fig:relaxation_and_cutoff_analysis}d-i. Starting from the Orbach mechanism, computed at 100 K for this analysis, we observe in Fig.~\ref{fig:relaxation_and_cutoff_analysis}d-f a sharp change in the relaxation time as the upper bound $\omega_\mathrm{c}$ of the energy window approaches the energy gap between the ground state and the first excited KD. For \textbf{1}, \textbf{2} and \textbf{3} this drop in $T_1$ is registered around values closely matching the first KD excitation. %431 cm$^{-1}$, 464 cm$^{-1}$ and 378 cm$^{-1}$, respectively. 
This behaviour highlights the key role of the initial absorption of a phonon that leads to a real spin transition from the ground state to the first excited KD. 
Conversely, the behaviour of $T_1$ upon increasing the lower bound $\omega_\mathrm{c}$ of the phonon energy window reveals an approximately specular trend, with an almost complete quenching of relaxation when phonon modes with energies around the first excited KD are excluded. Turning to the Raman regime, computed at 10 K (see Fig.~\ref{fig:relaxation_and_cutoff_analysis}f-i), the high energy cut-off analysis shows a rapid decay of $T_1$ 
upon inclusion of phonons within the first 30–100 cm$^{-1}$ for all three systems, demonstrating the role of low-energy phonons in Raman relaxation. Convergence is reached at $\approx$ 70 cm$^{-1}$ for \textbf{1}, $\approx$ 90 cm$^{-1}$ for \textbf{2} and $\approx$ 60 cm$^{-1}$ for \textbf{3}. The low-energy cut-off analysis shows that an almost complete quenching of the relaxation is reached only after the exclusion of the phonon modes up to 200\,cm$^{-1}$. This indicates that a broad set of vibrational modes could provide a contribution to the Raman relaxation process that is only slightly less important than the dominating low-energy phonons.

\begin{figure*}[ht]
    \centering    
    % --- Prima riga ---
    \begin{subfigure}{0.4\textwidth}
        \centering
        \includegraphics[width=\textwidth]{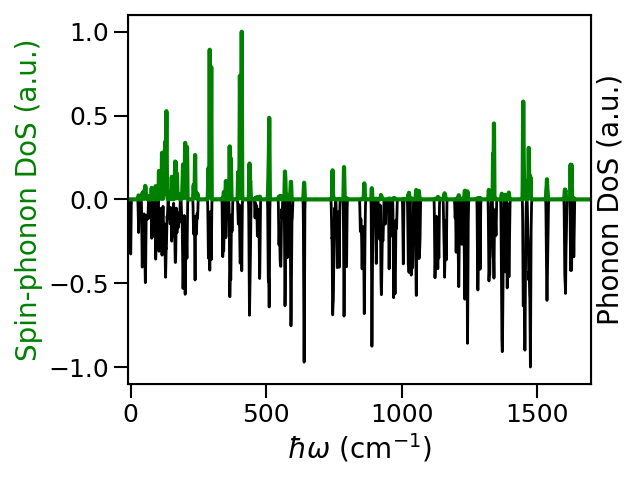}
    \caption{}
    \end{subfigure}
    \hspace{0.5cm}
    \begin{subfigure}{0.4\textwidth}
        \centering
        \includegraphics[width=\textwidth]{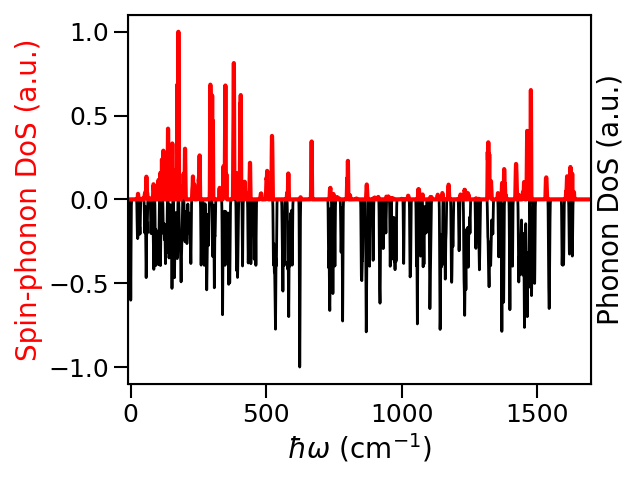}
        \caption{}
    \end{subfigure}
    % --- Seconda riga ---
        \begin{subfigure}{0.4\textwidth}
        \centering
        \includegraphics[width=\textwidth]{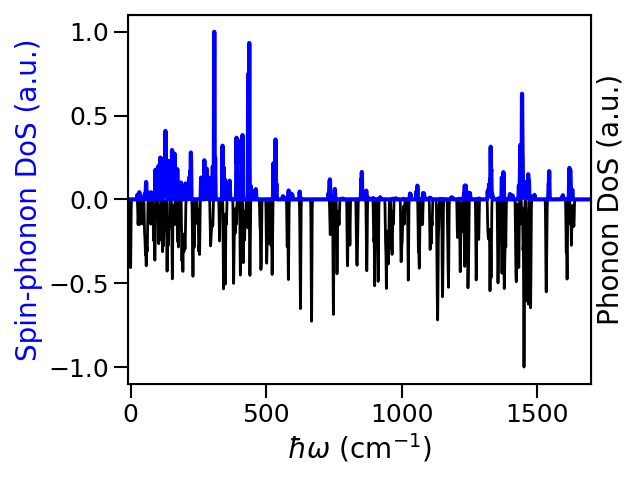}
        \caption{}
    \end{subfigure}
    \hspace{0.5cm}
    \begin{subfigure}{0.4\textwidth}
        \centering
        \includegraphics[width=\textwidth]{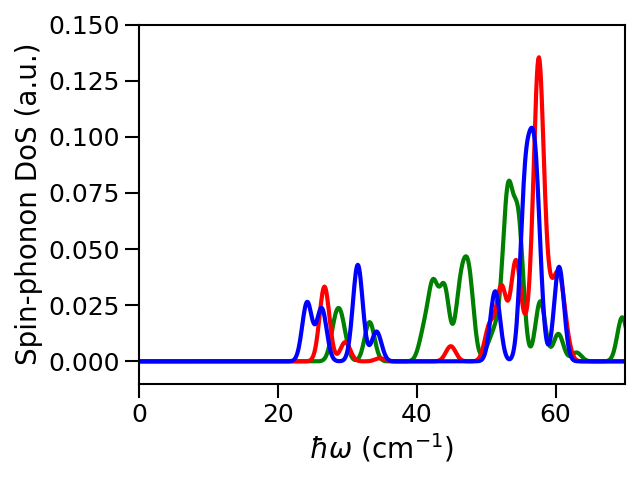}
        \caption{}
    \end{subfigure}
    % --- Terza riga ---
        \begin{subfigure}{\textwidth}
        \centering
        \includegraphics[width=0.30\textwidth]{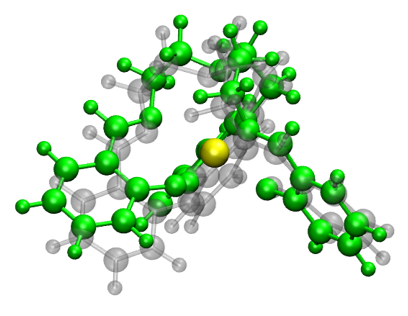}%
        \hfill
        \includegraphics[width=0.36\textwidth]{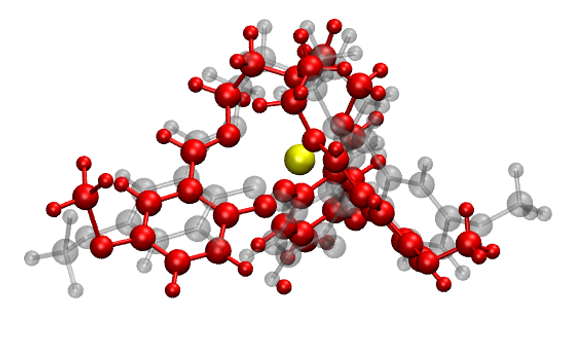}%
        \hfill
        \includegraphics[width=0.30\textwidth]{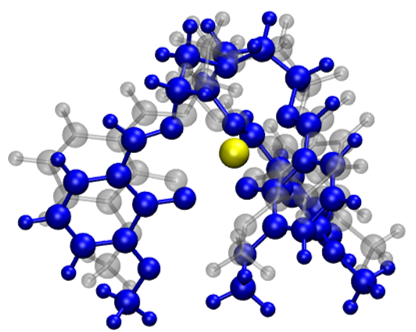}%
        \caption{}
    \end{subfigure}    
    \caption{Panels a-d: Normalized spin-phonon density of states for Yb(trensal) (a), Yb(trenpvan) (b) and Yb(trenovan) (c). In black: normalized phonon density of states for each system. Every spectrum is rescaled with respect to its maximum value (in modulus). The scale factor is about 0.91 for Yb(trensal), 0.55 for Yb(trenpvan) and 0.75 for Yb(trenovan). (d): zoom on the [0,70] cm$^{-1}$ interval. The adopted smearing is Gaussian and equal to 1 cm$^{-1}$. Panel e: Pictorial visualization of the low-energy oscillation modes, corresponding to the highest peaks of the spin-phonon DoS reported in (d), appearing at 53.3 cm$^{-1}$ (green), 56.6 cm$^{-1}$ (blue), and 57.6 cm$^{-1}$ (red), respectively. The geometry of the molecule at a certain instant (coloured molecule) is superimposed to the one at the beginning of the oscillation (shaded molecule). In this portrayal, the Yb atom colour is yellow to simplify its visualization.}
    \label{fig:pDOS-sphDOS}
\end{figure*}

As a final step, in order to identify the nature of the main vibrational modes that contribute to relaxation, we examine the spin-phonon coupling density of states $D(\omega)$, that is defined as:
\begin{equation}
    D(\omega)=\sum_\alpha \sum_{lm} \left( \frac{\partial B_m^l}{\partial Q_\alpha} \right)^2\delta(\omega-\omega_\alpha).
\end{equation}
The results are reported in Fig.~\ref{fig:pDOS-sphDOS}a-d, where we show both the normalized phonon density of states (black) and the quantity $D(\omega)$ rescaled with respect to its maximum value.
These results show that the minor chemical differences among the three variants affect the relaxation dynamics in a subtle manner. For such localized structural modifications, phonon modes are expected to be primarily altered in the high-energy part of the spectrum (hundreds of cm$^{-1}$), and indeed changes in this region were detected by IR spectroscopy \cite{Hansen2024a}. However, Fig.~\ref{fig:pDOS-sphDOS} reveals that the spin–phonon density of states of the three variants does not differ by a simple shift or rescaling, nor by the contribution of isolated phonon modes, but rather through an extended redistribution across the spectrum. At temperatures of the order of 10 K or below, our cut-off analysis shows that Raman relaxation is governed exclusively by phonons in the very low-energy range. The corresponding modes cannot be ascribed to the vibrations of a specific, localized group of atoms in the molecules. Indeed, a detailed inspection of the atomic displacements related to these modes clearly shows that the oscillations pertain to the molecular structure as a whole. We illustrate these findings in Fig. \ref{fig:pDOS-sphDOS}e, where we report the geometry of the molecule distorted according to the nature of low-energy phonons.      
\section*{Discussion and Conclusions}
\label{sec:conclusions}
We presented a comprehensive ab initio study of the relaxation behavior of a family  of virtually isostructural Yb(III) coordination compounds: Yb(trensal) and its derivatives Yb(trenpvan) and Yb(trenovan). The derivatives differ only by the substitution of a hydrogen atom with a methoxy group at distinct positions on the trensal backbone, resulting in minimal structural perturbations, yet sizable effects on spin relaxation. Multireference electronic structure calculations, in agreement with available experimental data, reveal nearly identical electronic excitations across the series, coherently with a high degree of similarity in their first coordination shells. This exceptional degree of electronic similarity makes these compounds an ideal platform to isolate and quantify the role of phonons in spin-lattice relaxation dynamics, which, unlike crystal field splitting, is expected to be influenced by the full molecular structure, beyond the first coordination shell, and the molecular crystal packing.

As expected from the magnitude of the crystal-field gaps, Orbach relaxation is strongly suppressed at low temperatures, leaving Raman relaxation as the dominant mechanism. The simulated relaxation times are consistent with experimental data over the temperature range where Raman relaxation prevails, while additional mechanisms become relevant below a few degrees Kelvin, such as quantum tunnelling of the magnetization \cite{Aravena2018,Mattioni2024}, or dipole–dipole interactions between electronic spins. In our calculations we assume perfectly diluted crystals of Yb(tren-) but in practice, even at 1$\%$ Yb concentration, such interactions can affect the relaxation dynamics at low temperatures, reducing $T_1$ and contributing to deviations from the simulated values. A comprehensive set of measurements at higher temperatures would therefore be highly valuable to further assess the agreement between our model and the experimental data. Nonetheless, our energy cut-off analysis shows that for $T \sim 10$ K, only phonons within a low-energy window of $\sim 60$ cm$^{-1}$ are relevant. At first glance, the minor structural differences among the variants, together with their nearly identical electronic spectra, would suggest an essentially unchanged low-energy spin-phonon coupling. Indeed, the structural perturbations are highly localized and spatially distant from the Yb ions, and would therefore be expected to primarily affect high-energy vibrational modes. Contrary to this expectation, Fig.~\ref{fig:pDOS-sphDOS} reveals marked differences in $D(\omega)$ already in the low-energy regime. These results demonstrate that even minimal structural changes can significantly modify low-energy spin–phonon coupling, leading to non-trivial variations in relaxation rates that originate from collective changes across the low-energy phonon manifold rather than from individual vibrational modes. These findings highlight the high sensitivity of low-temperature spin dynamics to weak spin–phonon interactions and show that even subtle chemical modifications in the second coordination sphere provide an effective handle to tune relaxation pathways. 

Importantly, these observations are at odds with the conventional approach to the chemical design of magnetic molecules. Until recently, the principal route toward the design of magnetic molecules has been the tailoring of crystal field excitations through the modification of the ions' first coordination shell. The overwhelming success of this strategy could potentially reinforce the idea that spin-phonon coupling could also be tackled this way, and attempts to control and rationalize Raman relaxation through a tailoring of molecular structure have been made. However, the present results demonstrate that this conceptual approach to the chemical design of magnetic molecules is no longer applicable when considering the problem of tuning spin-phonon coupling. Different from the well understood correlation between an ion's crystal field and first-coordination shell modification, even localized chemical perturbations at the structural level result in delocalized and non-trivial modifications of vibrational spectra and spin-phonon couplings to a degree that it is not possible to establish a clear cause-and-effect relation. This state of affairs clearly represents a challenge for the entire community, with decades-long consolidated approaches to the synthesis of molecular magnets in need of an update. At the same time, we believe that these outstanding challenges also constitute an opportunity for further development. The establishment of a systematic understanding of THz vibrations and their relation to molecular and crystal structure would have a broad impact on many different branches of chemistry \cite{Banks2023}, physics and materials science as well as positive repercussions of technologies ranging from heat transport \cite{Biswas2026} to vibrational spectroscopic imaging \cite{Cheng2015}. A range of tools is already available to pursue this challenge. On the one hand, first-principles methods like the ones used in this work are nowadays capable of providing quantitative predictions of relaxation times across a wide range of temperatures and stand out as a mature tool to address the correlation between spin-phonon and molecular structure. Recently, these tools have also been accelerated through machine learning surrogate models \cite{Briganti2025} as well as integrated into molecular discovery frameworks \cite{Frangoulis2025,Khatibi2026,Mariano2024a}, and further development in these directions holds a transformative potential for the field of molecular magnetism. On the other hand, a range of scattering and spectroscopic techniques have recently been used to support theoretical explorations of spin-phonon coupling, e.g. neutron or x-ray scattering \cite{Garlatti2020,Garlatti2023,Dunstan2023}, far-infrared or Raman magnetospectroscopy \cite{Kragskow2022,Moseley2018,Moseley2020}, nuclear resonance vibrational spectroscopy \cite{Scherthan2020}, and THz electron paramagnetic resonance \cite{Chen2026}. These techniques are increasingly successful in probing the spin-phonon coupling of phonons in resonance to spin transitions, and the exploration of their remit in sensing those low-energy phonons leading to virtual transitions and Raman relaxation appears as a very promising possible future direction for the field to take.

In conclusion, we have here used three Yb(III) compounds to isolate the correlation between molecular structure modifications and spin-phonon coupling and relaxation, going beyond well-documented first-coordination-shell effects on crystal field excitations. The absence of a simple chemical interpretation to our quantitaive simulations highlights the urgency to extend the existing frameworks for the molecular design of magnetic molecules beyond simple magneto-structural correlations and towards building a mapping between molecular structure and delocalized THz phonons and corresponding spin-phonon couplings. We anticipate that further development and integration of chemical synthesis, spectroscopical methods, and predictive first principles frameworks hold the key to solving this challenge.

\noindent
\\

\textbf{Acknowledgements and Funding} This project has received funding from the European Research Council (ERC) under the European Union’s Horizon 2020 research and innovation programme (grant agreement No. [948493]). Computational resources were provided by the Trinity College Research IT and the Irish Centre for High-End Computing (ICHEC). G.S., S.C. and P.S. acknowledge the European Union Next Generation EU, PNRR MUR projectPE0000023-NQSTI for financial support.

\bibliography{refs}

\end{document}